\documentstyle[preprint,aps,eqsecnum,fixes,epsfig,amstex]{revtex}

\makeatletter           
\@floatstrue
\def\figure{\let\@capwidth\columnwidth\@float{figure}}
\let\endfigure\end@float
\@namedef{figure*}{\let\@capwidth\textwidth\@dblfloat{figure}}
\@namedef{endfigure*}{\end@dblfloat}
\makeatother

\begin {document}

\tightenlines
\newcommand{\nc}{\newcommand}
\nc{\bea}{\begin{eqnarray}}
\nc{\eea}{\end{eqnarray}}
\nc{\beqa}{\begin{eqnarray}}
\nc{\eeqa}{\end{eqnarray}}
\nc{\NCS}{N_{\rm CS}}
\nc{\lsim}{\mbox{\raisebox{-.6ex}{~$\stackrel{<}{\sim}$~}}}
\nc{\gsim}{\mbox{\raisebox{-.6ex}{~$\stackrel{>}{\sim}$~}}}
\nc{\mD}{m_{\rm D}}
\nc{\meq}{m_{\rm eq}}
\nc{\mth}{m_{\rm th}}
\nc{\Teq}{T_{\rm eq}}
\hyphenation{de-scrip-tion}


\preprint {UW/PT 00-04}

\title{Electroweak Bubble Wall Friction:  Analytic Results} 

\author {Guy D. Moore}

\address
    {%
Department of Physics,
University of Washington, 
Seattle WA 98195-1560 USA	
    }%
\date {January 2000}

\maketitle
\vskip -20pt

\begin {abstract}%
{%
We present an entirely analytic, leading log order determination of the
friction an electroweak bubble wall feels during a first order
electroweak phase transition.  The friction is dominated by $W$ bosons,
and gives a wall velocity parametrically $\sim \alpha_w$, and
numerically small, $\sim .01 -\!\! \! \! - \: 0.1$ 
depending on the Higgs mass.
}%
\end {abstract}

\thispagestyle{empty}


\section {Introduction}

Electroweak baryogenesis is the name for the production of the baryon
number asymmetry of the universe at the electroweak epoch.  It is
possible in extensions of the standard model where the three famous
Sakharov conditions can be met:
\begin{enumerate}
\item Baryon number violation is efficient,
\item The discrete symmetries C and CP are violated, and
\item There is a departure from equilibrium, coincident with baryon
number violation turning off.
\end{enumerate}
All three of these criteria potentially exist in the standard model.
Baryon number violation is not only present, but efficient
\cite{Kuzmin,ArnoldMcLerran}.  The expansion of 
the universe provides departure from equilibrium.  And of course, C and
CP are known not to be true symmetries.

Whether or not electroweak baryogenesis can explain the size of the
observed baryon number of the universe, which is about \cite{Olive}
\begin{equation}
\frac{{\rm baryons}}{\rm entropy} \simeq ( 2 - \!\!\! - \: 7 ) 
	\times 10^{-11} \, ,
\end{equation}
is a more detailed question.  In the minimal standard model, both the
available CP violation \cite{Gavela}, and the departure from
equilibrium \cite{KLRSresults}, appear to be grossly insufficient.
However the question is quite open in extensions of the standard model.

Baryogenesis at a first order electroweak phase transition is a
complicated process,
and requires understanding several things.  First, we must be able to
compute the strength of the phase transition.  The tools for doing this
are now well developed \cite{KLRSresults,LaineRummukainen}.  Next, we
must be able to compute the efficiency of baryon number violation; here
the tools are also well developed
\cite{ASY,broken_nonpert,Bodeker,bodek_paper,BMR,AY2_short}.  
Finally one must be
able to compute the microscopic dynamics of baryon number carrying
excitations in the presence of an electroweak phase interface,
henceforth called the ``bubble wall''
\cite{baryo_refs}.  This problem is not well under
control, but the papers quoted show that the results strongly depend on
another factor, which is the velocity of propagation of the bubble
wall.  This paper will discuss the computation of this bubble wall
velocity, which is important for baryogenesis, and is 
also interesting as an 
example where such a dynamical quantity can be computed from first
principles.  

A fairly substantial literature already exists on the electroweak bubble 
wall velocity.  Since a bubble wall liberates latent heat as it
propagates, hydrodynamic considerations are potentially important.  
Hydrodynamic considerations appear in
\cite{old_hydro,Laine2,Heckler,stability,LaineKSuonio}.  The conclusion
is that, if the friction on the bubble wall is small, then the
hydrodynamics are important; but if the bubble friction is large, so the 
wall velocity is small, then all that matters is the general rise in
temperature from the bubbles in aggregate as the transition proceeds.

There is also a literature on the friction the bubble wall feels
\cite{Turok,Dine,McTurok,Khleb_wall,Arnold_wall,MooreProkopec,MooreTurok}.
The paper of Khlebnikov \cite{Khleb_wall} shows how the friction is
related to the self-energy of the zero mode of the bubble wall.  That is,
the friction arises from the back-reaction on the wall of the
disturbance from equilibrium of excitations, induced by the motion of
the wall.  Most of the papers quoted study this back reaction by
treating the excitations with kinetic theory.  The exception is
\cite{MooreTurok}, where it is argued that infrared SU(2) gauge field
and Higgs field excitations are
most important, and that it is more appropriate to treat them as
classical fields, which can be done nonperturbatively on the lattice.

Both the kinetic descriptions, and classical nonperturbative treatment,
missed one important piece of physics, however, which as we will see
leads them to be incorrect {\em parametrically}.  That is, they miss the 
physics of screening and Landau damping, which dominates the dynamics of 
infrared gauge fields.  The importance of this physics has been pointed
out by Arnold, Son, and Yaffe, \cite{ASY,HuetSon,Son}, in the context of 
determining the baryon number violation rate, and has been further
discussed in \cite{Bodeker,moreBodeker,moreASY,AY2_short}.  The
central result is that the SU(2) gauge field $A$, 
instead of evolving under (classical) equations
of motion of the form (in temporal gauge, ignoring nonlinearities)
\begin{equation}
\frac{d^2 A}{dt^2} = - (k^2 + m^2) \, A \, ,
\end{equation}
with $m^2 \! = \! (g^2/4)\, \phi^2$ the mass squared induced by a Higgs
condensate, instead undergoes {\em overdamped} evolution,
\begin{equation}
\label{overdampedA}
\frac{\pi \mD^2}{4 k} \: \frac{dA}{dt} = -(k^2 + m^2) \, A 
	+ {\rm noise} \, ,
\end{equation}
with $\mD^2$ the Debye mass squared, $\mD^2 = (11/6)\, g^2T^2$ in the
standard model.
In this letter we will see what consequences this has for the kinetic
description of the bubble wall friction.

\section{Ingredients}

For concreteness we will work here with the minimal standard model, even 
though baryogenesis in that model is ruled out.  Since only the gauge
fields are overdamped, and since only the SU(2) and U(1) fields have
large interactions with the Higgs condensate, the extension to models
with more fields, such as the minimal supersymmetric standard model,
should be straightforward.  From here on we will also neglect the U(1)
field, which is the same as taking $g' \ll g$.  This is probably a
reasonable approximation.  Going beyond it would increase the
bookkeeping but would not make our treatment substantially more
difficult.

This work will be strictly analytic and
strictly based on parametric expansions in formally small quantities.
In order to obtain an electroweak phase transition of a strength which
can be analyzed perturbatively, we will take
\begin{equation}
g^3 \ll \lambda \ll g^2 / \log(1/g) \, ,
\end{equation}
where $\lambda$ is the Higgs self-coupling in the same normalization as
used in \cite{Dine}.  As we see in a moment, the gauge field condensate
is parametrically $\phi_0 \sim g^3T / \lambda$, so the induced gauge
field masses are $g \phi_0/2 \sim g^4T / \lambda$.  
The longitudinal gauge fields
have a Debye mass $\sim gT \gg g^4T/\lambda$ and can be neglected
\cite{Dine}; the one loop effective potential is then approximately
\cite{Dine}
\begin{equation}
\label{Veff}
V_{\rm 1\; loop}(\phi) = \frac{m^2(T)}{2} \phi^2 - -\frac{g^3T}{16 \pi} 
	\phi^3 + \frac{\lambda}{4} \phi^4 \, .
\end{equation}
A broken minimum exists if $dV/d\phi = 0$ at some nonzero $\phi_0$.  The 
broken minimum is degenerate with the minimum at $\phi=0$ at 
\begin{equation}
\label{Teq}
m^2(\Teq) = \frac{g^6 T^2}{128 \pi^2 \lambda} \, ,
\end{equation}
and the value of $\phi_0$ is
\begin{equation}
\phi_0(\Teq) = \frac{g^3 T}{8 \pi \lambda} \, .
\end{equation}
The $\phi$ profile of the electroweak bubble wall at $\Teq$ 
is given at leading order by 
\begin{equation}
\label{wall_shape}
\phi(z) = \frac{\phi_0}{2} \left[ 1 + \tanh \frac{z}{L} \right] \, , 
\qquad L = \frac{2}{m(\Teq)} \, .
\end{equation}
Here $z$ is a space coordinate orthogonal to the bubble wall.

The gauge field mass times the wall width,
\begin{equation}
\label{LmW}
L m_W = L \frac{g \phi}{2} = \frac{16 \pi \sqrt{2\lambda}}{g^3T} \,
	\frac{g^4T}{16\pi \lambda} \, \frac{\phi}{\phi_0}
	= \frac{\phi}{\phi_0} \, \sqrt{\frac{2g^2}{\lambda}} \gg 1 \, , 
\end{equation}
is large, and it is therefore possible to treat gauge field excitations, 
in the broken phase and inside the bubble wall, with kinetic theory.
The kinetic theory description, which amounts to taking the Higgs 
field background as approximately homogeneous and expanding in its
gradients, breaks down at the small $z$ (symmetric phase) edge
of the bubble wall, when $(\phi / \phi_0) \sim \sqrt{\lambda/g^2}$.
In fact, since $\lambda = (g^2/8)$ is the condition for $m_H = m_W$, the
kinetic description works fairly well at remarkably large Higgs masses.

All of the equations above are valid at leading order in $\lambda/g^2$
or $g^3/\lambda$.  when the former breaks down, higher loop corrections
and corrections from Higgs loops (neglected here) become important.
When the latter breaks down, the longitudinal gauge fields become
important.  We also need $g^3 \ll \lambda$ to ensure that the relevant
gauge fields will be overdamped.

\section{Friction}

The friction on an electroweak bubble wall is defined as the excess
pressure on the wall (directed towards the broken phase), 
over the equilibrium value;
\begin{equation}
{\rm friction} = P - P_{\rm eq} = P + V(\phi=0) - V(\phi_0) \equiv 
	P + \Delta V \, .
\end{equation}
We expect the friction to depend on the bubble wall velocity, with the
condition $P \! = \! 0$ determining the steady state bubble wall velocity
$v_w$, which is what we want to know.
We can define a linear response friction coefficient $\eta$ as a limit 
\begin{equation}
\eta \equiv \lim_{\Delta V \rightarrow 0} \frac{\Delta V}{v_w} =
	\frac{{\rm steady \; state \; friction}}{v_w} \, .
\end{equation}
This is what we want to determine.

The friction on the bubble wall depends on the departure from
equilibrium of the plasma excitations inside the bubble wall.  In a
kinetic theory description, the excitations are described by population
functions $f(k,x)$.  We write them as an equilibrium part $f_0$,
\begin{equation}
f_0 = \frac{1}{\exp(E/T) \pm 1} \, , \qquad E = \sqrt{k^2 + m^2(x)} \, ,
\end{equation}
with $+$ for fermions and $-$ for bosons, which is the case we will
care about, plus a departure from equilibrium $\delta f$.  The friction 
a bubble wall feels is, in the kinetic description
\cite{Turok,Dine,McTurok,MooreProkopec}\footnote{A derivation of sorts
can be found in \protect{\cite{MooreProkopec}}, but the expression is
implicit in the earlier references as well.},
\begin{equation}
\label{friction_is}
{\rm friction} = \int_{- \infty}^{\infty} \! \! dz 
	\sum_{\rm DOF}
	\int \! \frac{d^3k}{(2\pi)^3} \frac{d\phi}{dz} \:
	\frac{dm^2}{d\phi} \: \frac{dE}{dm^2} \: \delta f \, ,
\end{equation}
where the sum is over degrees of freedom which get a mass from the Higgs 
field.
This expression has a clear intuitive meaning; it is the sum over excess 
particles of $(dE/dz)$, the force the wall exerts on them.
It remains to determine $\delta f$ and evaluate the integral.  

The friction will be dominated by the gauge boson contribution.  The
contribution from Higgs bosons is smaller because their $dm^2/d\phi$ is
smaller by $\lambda / g^2$,
and because their evolution is not overdamped.  The
fermionic contributions, such as that from the top quark, are
smaller because Fermi-Dirac statistics lack the infrared divergence of
Bose-Einstein statistics, so their $\delta f$ has much weaker infrared
behavior.  The parametric argument appears in
\cite{MooreTurok}, and a reasonable estimate of friction from top quarks 
appears in \cite{MooreProkopec}; it proves numerically smaller than what 
we find below.  The case where there is a light scalar top is more difficult
and we do not consider it; for the MSSM the friction we find should be
viewed as a lower bound rather than a tight estimate.

As discussed above, the gauge fields undergo overdamped evolution given
by Eq.~(\ref{overdampedA}).  Since $f \propto A^2$, the equation for $f$ 
is
\begin{equation}
\frac{\pi \mD^2}{8k} \: \frac{df}{dt} = - E^2 f 
	+ {\rm noise} \, ,
\end{equation}
where the noise is of the right size to ensure that, for $m^2$ time
independent, $f$ will approach $f_0$; so averaging over the noise,
\begin{equation}
(k^2+m^2) f + {\rm noise} \rightarrow E^2 \delta f \, .
\end{equation}
Also, $df/dt = df_0/dt + d (\delta f)/dt$.  At small $v_w$, the limit we 
are interested in, $\delta f \ll f_0$, and $d(\delta f)/dt$ may be dropped.
Further, 
\begin{equation}
\frac{df_0}{dt} = \frac{d\phi}{dt} \frac{dm^2}{d\phi} 
	\frac{dE}{dm^2} \frac{df_0}{dE} = - v_w \frac{d\phi}{dz} 
	\frac{dm^2}{d\phi} \frac{1}{2ET} f_0(1 + f_0) \, .
\end{equation}
Therefore, the departure from equilibrium of a gauge boson degree of
freedom is
\begin{equation}
\label{delta_f}
\delta f = \frac{\pi \mD^2 v_w}{16 k E^3 T} f_0(1+f_0) 
	\frac{d\phi}{dz}\frac{dm^2_W}{d\phi} \, .
\end{equation}
Note that transport plays no role in setting $\delta f$; this is because 
the gauge fields are overdamped.
Substituting this into Eq.~(\ref{friction_is}), and noting that there
are 6 species of transverse $W$ bosons (3 flavors times 2 spins), gives
\begin{equation}
\label{friction2}
{\rm friction} = \frac{6 \pi v_w \mD^2}{8} \int_{-\infty}^{\infty} 
	\! \! \! \! dz \left( \frac{d\phi}{dz} \: \frac{dm_W^2}{d\phi} 
	\right)^2 
	\int \frac{d^3k}{(2 \pi)^3} \frac{f_0 (1+f_0)}{4kE^4T} \, .
\end{equation}
Since $f_0$ is monotonically decreasing, the momentum integral is
infrared dominated, cut off by the nonvanishing $W$ boson mass.
Therefore it is appropriate to make the approximation, valid in the
infrared, that $f_0 \simeq 1+f_0 \simeq T/E$, and evaluate the integral;
\begin{equation} \hspace{-0.5in}
\int \! \frac{d^3 k}{(2\pi)^3} \frac{f_0 (1+f_0)}{4kE^4T} \simeq
	\int_0^{\infty} \frac{k^2 dk}{2\pi^2} \frac{T}{4kE^6} = 
	\frac{T}{16 \pi^2} \int_{m_W^2}^\infty \frac{d(E^2)}{E^6}
	= \frac{T}{32 \pi^2 m_W^4} \, . \hspace{-0.5in}
\end{equation}
The friction, using $m_W = g \phi/2$, is then 
\begin{equation}
\label{friction3}
{\rm friction} = v_w \frac{3 \mD^2 T}{32 \pi} \int_{-\infty}^\infty 
	dz \left( \frac{d\phi}{dz} \right)^2 \frac{1}{\phi^2} \, .
\end{equation}
Writing $(d\phi/dz)dz = d\phi$, and using Eq.~(\ref{wall_shape}) to
write
\begin{equation}
\frac{d\phi}{dz} = \frac{2\phi (\phi_0 - \phi)}{L\phi_0} \, ,
\end{equation}
Eq.~(\ref{friction3}) now gives
\begin{equation}
\label{friction4}
\eta = \frac{3 \mD^2 T}{16 \pi L} \times \left( 
	\int_0^{\phi_0} \frac{(\phi_0 - \phi) d\phi}{\phi_0 \phi}
	= \int_0^1 \frac{(1-x) dx}{x} \right) \, .
\end{equation}
There is a log divergence arising from the symmetric phase side of the
bubble wall.  The log will be cut off where the first approximation used 
to derive Eq.~(\ref{friction3}) breaks down.  The perturbative expansion 
breaks down when $m_W \sim g^2 T$, or $\phi \sim gT$, which is at
$(\phi/\phi_0) = (\lambda/g^2)$.  The kinetic theory description breaks
down at $(\phi/\phi_0) = \sqrt{\lambda/g^2}$, see Eq.~(\ref{LmW}); this
occurs first.  

Since the degrees of freedom which dominate the friction
are those with $k \sim m$, when $m$ drops below $1/L$, it is no longer
appropriate to treat the particles as seeing a slowly varying wall.
Such degrees of freedom see a wall which is sharper than their
wavelength can resolve.  For those degrees of freedom with $kL \gg 1$,
we find the friction scales as $1/L$.  This must go over to an $L$
independent value for wavelengths which cannot resolve the thickness of
the wall, which means that their contribution is less than the kinetic
theory estimate.  Hence the log is cut off at $(\phi/\phi_0) \sim
\sqrt{\lambda / g^2}$, the contribution from very infrared degrees of
freedom is subdominant.  Hence the friction we determine is
\begin{equation}
\label{friction5}
\hspace{-0.7in} \eta = \frac{3 \mD^2 T}{16 \pi L} \Big( \log 
	(m_W L) + O(1) \Big)
	= \frac{3}{16 \sqrt{2}}\; \frac{m_D^2}{g^2T^2}\;
	\frac{g}{\sqrt{\lambda}}\;
	\alpha_w^2 T^4 \left( \log \frac{g}{\sqrt{\lambda}} 
	+ O(1) \right) \, .\hspace{-0.4in}
\end{equation}
The first expression makes no assumptions about the effective potential
or wall thickness and should be valid in extensions as well as the
standard model.

Now we comment on the parametric form of $\eta$.
Taking $\lambda \sim g^2$ and neglecting logs, 
the friction coefficient is $\eta \propto \alpha_w^2
T^4$.  We can guess this on dimensional grounds by noting that
\begin{equation}
\big[ \eta \big] = \left[ \frac{{\rm pressure}}{{\rm velocity}} 
	\right] = \left[
	\frac{{\rm energy} \times {\rm time}}{{\rm length}^4} \right] \, ,
\end{equation}
and that $\eta$ arises from infrared gauge field physics.  Such physics
has a natural energy scale $\sim T$, a natural length scale 
$\sim 1 / \alpha_w T$, and a natural time scale $\sim 1 / (\alpha_w^2 T)$
\cite{ASY}; so on dimensional grounds we should have anticipated $\eta
\sim \alpha_w^2 T^4$.  This is to be contrasted with the pressure
driving the bubble wall, which by the same parametric estimates must be 
\begin{equation}
\big[ P \, \big] = \left[ \frac{{\rm energy}}{{\rm length}^3} \right]
	\, , \qquad P \sim \alpha_w^3 T^4 \, .
\end{equation}
Hence, the bubble wall velocity is parametrically $v_w \sim \alpha_w$.

\section{B\"{o}deker's Effective Theory}

The reason we considered the case $\lambda \ll g^2 / \log(1/g)$, rather
than $\lambda \ll g^2$, is because in the parametric regime
\begin{equation}
\frac{g^2}{\log(1/g)} \ll \lambda \ll g^2
\end{equation}
The gauge bosons with $m \sim g \phi_0/2$ do not obey
Eq.~(\ref{overdampedA}); instead B\"{o}deker's effective theory is
applicable \cite{Bodeker};
\begin{equation}
\sigma \frac{dA}{dt} = - E^2 A + {\rm noise} \, , \qquad
\sigma = \frac{2 \pi \mD^2}{3 g^2T \log(1/g)} \, ,
\end{equation}
up to corrections suppressed by $\log(1/g)$.  In this case the
derivation proceeds analogously, but the behavior is slightly less
infrared dominated, and no log occurs in the $\phi$ integral.  
The final expression is
\begin{equation}
\eta = \frac{1}{256 \sqrt{2}\: \log(1/g)} \left( \frac{g}{\sqrt{\lambda}} 
	\right)^3 \left( \frac{m_D^2}{g^2T^2} \right) 
	\alpha_w^2 T^4 \, .
\end{equation}
This result is in parametric agreement with what we found above.  It is
only applicable in an extremely narrow parametric range, and receives
corrections suppressed only by $(\lambda/g^2)$ (from the loop expansion) 
and $g^2 \log(1/g) / \lambda$ (from the breakdown of B\"{o}deker's
effective theory).

\section{Conclusion:  Value of the Wall Velocity}

Since we only have the friction at leading log, which means with at least
a factor of $2$ error, we will do with a fairly crude estimate of the
pressure which drives the bubble.  In references \cite{Dine,McTurok} the 
nucleation temperature is estimated as occurring at $m^2(T_{\rm nuc}) =
0.8 m^2(\Teq)$.  For this value, using Eqs. (\ref{Veff}) and (\ref{Teq}), 
the pressure driving the bubble wall is 
approximately $P = .94 \times (g^6/1024 \pi \lambda^3) \alpha_w^3 T^4$,
and the wall velocity is about
\begin{equation}
v_w  = \frac{P}{\eta} \simeq .0012 \left( \frac{g}{\sqrt{\lambda}}
	\right)^5 \frac{\alpha_w}{\log(g/\sqrt{\lambda})} \, ,
\end{equation}
which, for $m_H \simeq m_W$ or $\lambda = g^2/8$, and estimating $\log
(g / \sqrt{\lambda}) \simeq 1$, is $v_w \simeq \alpha_w/4 < 0.01$.  For
$\lambda / g^2 \simeq .04$, as required to give a sufficiently strong
phase transition, $v_w \sim 0.1$.  The
numerical value for the bubble wall velocity is expected to be small.
The numerical estimate from the result determined using B\"{o}deker's
effective theory is similar; $v_w \sim 0.2$ for $\lambda / g^2 = .04$.  
Neither estimate is 
very reliable because the constant under the log has not been
determined; but the conclusion $v_w \ll 1$ is clear.

The friction we find is larger, and the value of $v_w$ smaller, than in
previous literature.  In particular we find a much larger friction than
the numerical results of \cite{MooreTurok} indicate.  This is partly
because there we did not include hard thermal loop effects for the gauge 
fields, but overly aggressive data fitting may have contributed.
It would be interesting to make a new numerical analysis using the
techniques developed in \cite{BMR}.

We should also briefly comment on how the result may change in
extensions to the standard model and beyond leading order in $v_w$.
Beyond first order in $v_w$, a few effects become important.  We may not 
be able to neglect $\delta f$ next to $f_0$ in getting
Eq.~(\ref{delta_f}); $\delta f$, and the friction, will be larger.
Also, the frictive pressure will change the bubble wall shape.  Since
most of the friction is on the symmetric phase side of the wall, the
wall will become narrower, see \cite{MooreProkopec}, which also
increases the friction.  Finally, in extensions to the standard model
(which are the only viable candidates for baryogenesis because they can
provide both a strong phase transition and the heavy Higgs boson
required by experiment), the bubble wall is typically thinner, because
the Higgs mass is larger.  This increases the friction, even before
considering new contributions from extra light bosons such as a light
scalar top.  For these reasons we anticipate that electroweak bubble
wall velocities are quite generally much less than 1.

\end{document}